\def\isaccepted{1}

\documentclass[conference]{IEEEtran}
\ifCLASSINFOpdf
   \usepackage[pdftex]{graphicx}
   \DeclareGraphicsExtensions{.pdf,.jpeg,.png}
\else
\fi
\usepackage{tabularx,calc}
\usepackage{color}
\usepackage{tubscolors}
\usepackage{tikz}
\usetikzlibrary{shapes,automata,arrows,matrix,backgrounds,fit,patterns,decorations.markings, svg.path, shapes.multipart, external, shadows, positioning, calc, decorations}
\usepackage{siunitx}
\usepackage[style=ieee,backend=bibtex,minnames=1,maxcitenames=2,maxbibnames=99,doi=false,isbn=false,url=false,natbib=true]{biblatex} 
\usepackage[caption=false, font=footnotesize]{subfig}
\usepackage{ifthen}
\makeatletter
\newcounter{IEEE@bibentries}
\renewcommand\IEEEtriggeratref[1]{%
	\renewbibmacro{finentry}{%
		\stepcounter{IEEE@bibentries}%
		\ifthenelse{\equal{\value{IEEE@bibentries}}{#1}}
		{\finentry\@IEEEtriggercmd}
		{\finentry}%
	}%
}
\makeatother

\newcommand\copyrighttext{%
	\footnotesize%
	 \parbox[t]{.1\textwidth}{\copyright{} \the\year~IEEE.}%
	 \parbox[t]{.9\textwidth}{%
	 	Personal use of this material is permitted. Permission from IEEE must be obtained for all other uses, in any current or future media, including reprinting/republishing this material for advertising or promotional purposes, creating new collective works, for resale or redistribution to servers or lists, or reuse of any copyrighted component of this work in other works.%
 	 }%
}
\newcommand\copyrightnotice{%
	\begin{tikzpicture}[remember picture,overlay]
	\node[anchor=south,yshift=20pt] at (current page.south) {\parbox{\dimexpr\textwidth-\fboxsep-\fboxrule\relax}{\copyrighttext}};
	\end{tikzpicture}%
}

\tikzstyle{shadedBlue} = [top color=tuDarkBlue20, bottom color=tuDarkBlue40, draw=tuDarkBlue80, thick]%
\tikzstyle{shadedRed} = [top color=tuRed20, bottom color=tuRed40, draw=tuRed80, thick]%
\tikzstyle{shadedOrange}  = [top color=white, bottom color=tuOrange40,  draw=tuOrange100, thick]%
\tikzstyle{shadedGreen}  = [top color=white, bottom color=tuGreen40,  draw=tuGreen100, thick]%
\tikzstyle{shadedYellow} = [top color=white, bottom color=tuYellow40,  draw=tuYellow100, thick]%
\tikzstyle{shadedGray} = [top color=white, bottom color=tuGray20, draw=tuGray60, thick]%
\tikzstyle{shadedGrayLight} = [top color=tuBlack!5, bottom color=tuBlack!10, draw=tuBlack!30, thick]%
\tikzstyle{shadow} = [drop shadow={opacity=.5,shadow xshift=.3ex,shadow yshift=-.3ex}]%
\tikzstyle{triangle} = [isosceles triangle,isosceles triangle stretches]%
\tikzstyle{label-it} = [font=\itshape]%
\tikzstyle{block} = [draw, shadedBlue, rectangle, rounded corners, minimum height=2em, minimum width=5em]%
\tikzstyle{smallblock} = [draw, shadedBlue, rectangle, rounded corners, minimum height=1em, minimum width=2em,shadow]%
\tikzstyle{outerblock} = [draw, shadedGrayLight, draw=tuGray80, rectangle, rounded corners, minimum height=2em, minimum width=5em,shadow]%
\tikzstyle{bubble} = [fill=black,shadow,circle,draw=black,inner sep=0pt,minimum size=5pt]%
\tikzstyle{memory} = [cylinder, shape border rotate=90, aspect=.4, shadedGrayLight, minimum width=5em, shadow]%
\tikzstyle{inheritArrow} = [-open triangle 60,thick]%
\tikzstyle{kompArrow}    = [diamond-,thick]%
\tikzstyle{flowDecision} = [diamond, draw, shadedRed, text badly centered, inner sep=0pt,shadow]%
\tikzstyle{flowBlock} = [rectangle, draw, shadedBlue, text centered, rounded corners, minimum height=2em,shadow]%
\tikzstyle{bgBox} = [rectangle, draw, shadedGrayLight, text centered, rounded corners=3mm, shadow, inner sep=10pt]%
\tikzstyle{blockarrow} = [draw, thick, single arrow, minimum height=3em]%

\begin{document}
	
\title{Towards a Skill- And Ability-Based Development Process for Self-Aware Automated Road Vehicles}

\author{\IEEEauthorblockN{Marcus Nolte, Gerrit Bagschik, Inga Jatzkowski, Torben Stolte, Andreas Reschka and Markus Maurer}
\IEEEauthorblockA{Institute of Control Engineering\\
Technische Universit\"at Braunschweig\\
Braunschweig, Germany\\
Email: \{nolte, bagschik, jatzkowski, stolte, reschka, maurer\}@ifr.ing.tu-bs.de}
}
\maketitle

\newtheorem{definition}{Definition}

\begin{abstract}
	The development of fully automated vehicles imposes new challenges in the development process and during the operation of such vehicles.
	As traditional design methods are not sufficient to account for the huge variety of scenarios which will be encountered by (fully) automated vehicles, approaches for designing safe systems must be extended in order to allow for an ISO~26262 compliant development process.
	During operation of vehicles implementing SAE Levels 3+ safe behavior must always be guaranteed, as the human driver is not or not immediately available as a fall-back.
	Thus, the vehicle must be aware of its current performance and remaining abilities at all times.
	In this paper we combine insights from two research projects for showing how a skill- and ability-based approach can provide a basis for the development phase and operation of self-aware automated road vehicles.
\end{abstract}

\ifx \isaccepted \undefined
\else
\copyrightnotice
\fi

\section{Introduction}

The recent years have shown a growing trend towards the development of automated vehicles in the automotive industry.
Research institutions as well as original equipment manufacturers (OEMs) and startup companies are working towards the market introduction of vehicles which are capable of taking over the whole driving task without or with little supervision of the human driver.
Results from different research projects and already available partially automated driving systems (nomenclature according to SAE standard J3016 \cite{sae_j3016:_2016}) show potential customer benefit.
However, all of the recently presented automated vehicles driving on public roads (e.g.~\cite{tesla}, \cite{otto2016}) still require a human driver and/or external supervisor(s) to monitor the driving performance and to allow manual intervention if the vehicle behaves in an unsafe way.
As a result, current production systems have to be considered SAE Level 2 systems.
Until highly or even fully automated vehicles are ready for series deployment, manufacturers must ensure that the vehicles are safe in all regards of the complex driving task.
Thus, for higher levels of automation, the system must be aware of its performance limits in other words be \emph{self-aware} at the functional as well as on the technical level~\cite{reschka_conditions_2015, schlatow2017}.
Self-awareness means, that the vehicle must not only be aware of its external context (other vehicles, infrastructure, etc.), but also of its internal state in order to relate the vehicle's current performance to the system's performance limits.
When it comes to designing such systems, a development process is required, which provides means to seamlessly integrate the development of monitoring mechanisms.

Regarding the development process, the ISO~26262~standard defines the state-of-the-art for functional safety of E/E-systems in automotive applications~\cite{ISO_26262_2001}.
During the concept phase of the development process from the ISO~26262~standard, a hazard analysis and risk assessment (HARA) based on expert knowledge has to be conducted to identify and classify hazards originating from the considered system.
For fully automated vehicles, this is particularly challenging, as the category of controllability for the driver during risk assessment is not applicable, as the driver is not present in the vehicle's control loop.
Controllability for surrounding traffic participants is difficult to determine without a detailed scenario description.
From the resulting hazardous scenarios (definition according to~\cite{stolte2017}) safety goals have to be derived which yield top-level functional safety requirements.
Afterwards, requirements for functional components in the system must be identified and further detailed to derive technical safety measures.
Monitoring is a key enabler for automated vehicles and must supervise the system's adherence to the derived safety and technical requirements during operation.

This contribution will summarize recent activities at the Institute of Control Engineering of TU Braunschweig towards a specification of an ISO-26262-conforming concept phase for safe self-aware automated vehicles as well as strategies for runtime monitoring.
Examples from two current public research projects at the institute are utilized to describe the concept phase of a development process and how the results from the different process steps can be used to set up a framework for runtime monitoring and self-awareness. 

The project \emph{Controlling Concurrent Change} (CCC)\footnote{http://ccc-project.org/} aims at the design of model-based methods and mechanisms for runtime based integration processes.
One project goal is to develop a middleware which implements these methods.
At the same time it provides monitoring at different system levels which is required to supervise vehicle operation on the road.

The project \emph{Unmanned Protective Vehicle for Highway Hard Shoulder Road Works} (aFAS\footnote{This abbreviation is derived from the German project name.}) serves as an example for the design phase.
aFAS aims at developing an unmanned protective vehicle which is able to operate without human supervision on the hard shoulder (SAE Level~4).
A more detailed outline of the project is presented by Stolte et al.~\cite{stolte_towards_2015}.

\section{Project aFAS}
\label{ref:afas}
\begin{figure}[htb]
	\vspace{-0.5em}
	\centering%
	\scalebox{0.79}{%
		\usetikzlibrary{arrows}
\begin{tikzpicture}[x=1cm, y=1cm]
\definecolor{darkgrey} {RGB}{127,127,127}
\definecolor{lightgrey}{RGB}{229,229,229}

\draw [rounded corners, fill=lightgrey!50] (-1.5,0.55) rectangle +(7,-4.25);
\node [anchor = north west, font=\large] at (-1.5,0.55) {Vehicle Guidance System active};
\node [circle,minimum size=2cm, fill=darkgrey,align=center, text=white] (cm) at (0,-1) {Coupled \\Mode};
\node [circle,minimum size=2cm, fill=darkgrey,align=center, text=white] (fm) at (4,-1) {Follow \\Mode};
\node [circle,minimum size=2cm, fill=darkgrey,align=center, text=white] (sh) at (2,-2.5) {Safe \\Halt};
\node [circle,minimum size=2cm, fill=darkgrey,align=center, text=white] (mm) at (2,-5) {Manual\\Mode};
\node [circle,minimum size=0.75cm, fill=black,align=center] (start) at (0,-5) {};


\path [-latex, thick]				(start) edge (mm);
\path [-latex, thick]				(fm) edge (mm);
\path [-latex, thick]				(cm) edge (mm);
\path [-latex, thick, bend left=15] (mm) edge (sh);
\path [-latex, thick, bend left=15] (sh) edge (mm);
\path [-latex, thick, bend left=15] (cm) edge (sh);
\path [-latex, thick, bend left=15] (sh) edge (cm);
\path [-latex, thick, bend left=15] (fm) edge (sh);
\path [-latex, thick, bend left=15] (sh) edge (fm);

\end{tikzpicture}%
	}
	\vspace{-0.5em}%
	\caption{Operating modes of the unmanned protective vehicle}%
	\label{fig:statechart}%
	\vspace{-0.5em}%
\end{figure}
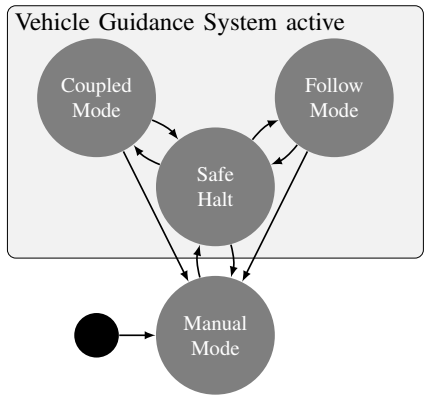

A major cause of fatalities of road workers in Germany are trucks crashing into the protective vehicle on the hard shoulder.
This is a main motivation for the development of an unmanned, fully automated protective vehicle.
In the project context, the consortium has produced results which are demanded by the concept phase of the ISO~26262~standard.
The concept of the protective vehicle includes four operating modes (cf.~Fig. \ref{fig:statechart}).

At the beginning of operation (black circle), the protective vehicle will be driven manually to the construction site.
From \emph{Manual Mode}, the operator can change the system's state to \emph{Safe Halt}.
\emph{Safe Halt} is considered the safe state of the protective vehicle.
This has proven to be an enabling factor of the project, as for automated vehicles with a wider range of functions, it is difficult to define a safe state for every possible scenario \cite{reschka_conditions_2015}.
Stopping is an applicable safety maneuver in any scenario for the planned use-case due to the low velocity and the limited area of operation.

In this operating mode the protective vehicle is slowed down to and eventually kept in standstill.
\emph{Safe Halt} also acts as the transition point into automated operating modes \emph{Coupled Mode} and \emph{Follow Mode}.
On- and off-ramps are passed in \emph{Coupled Mode} because they differ from normal operation on a hard shoulder and are more challenging for the vehicle's perception system.
In \emph{Coupled Mode} the protective vehicle is virtually coupled to the motion of the leading vehicle and follows its trajectory without consideration of lane markings.
This virtual drawbar is implemented by vehicle-to-vehicle communication between the leading and the protective vehicle.
The main operating mode for safeguarding on the hard shoulder is \emph{Follow Mode}.
In \emph{Follow Mode} the protective vehicle follows the left lane marking of the hard shoulder and keeps a distance to the leading vehicle of about 80 to \SI{100}{\meter}.
Due to regulations from UN-ECE~R79, the velocity of the protective vehicle is limited to \SI{10}{\kilo\meter/\hour}.

After having defined the functional range and the operational environment in the item definition, a HARA was conducted, as demanded in the ISO~26262~standard.
This yielded 59 hazardous scenarios and 17 resulting safety goals. 
As this contribution aims at proposing methods for different parts of a development process, we choose one safety goal for the \emph{Follow Mode} as a use-case.

One of the obvious hazards is the protective vehicle crossing (multiple) lanes on the highway and entering moving traffic.
This hazard can manifest in two different ways:
Table~\ref{tab:hazard} shows the corresponding safety goals which were chosen to mitigate (1) and eliminate (2) this hazard in the safety concept.
\begin{table}[h]
	\vspace{-1.2em}
	\caption{Hazardous events for protective vehicle entering traffic, related safety goals and Automotive Safety Integrity Levels.}
	\vspace{-0.25em}
	\renewcommand{\arraystretch}{1.25}
	\centering
	\begin{tabularx}{0.49\textwidth}{|>{\hsize=0.5\hsize}X>{\hsize=0.4\hsize}X>{\hsize=0.1\hsize}X|}
	\hline 
	\textbf{Hazardous Event} & \textbf{Safety Goal} & \textbf{ASIL} \\ 
	\hline 
	(1) Protective vehicle enters & Limit left lock & D \\ 
	moving traffic with full & to \SI{3}{\degree}. & \\
	left lock (maximal steering angle).  & & \\
	\hline 
	(2) Protective vehicle enters & Always maintain a & B \\ 
	moving traffic with limited & safe distance to & \\
	left lock. & the left lane marking. & \\
	\hline 
	\end{tabularx} 
	\label{tab:hazard}
	\vspace{-0.6em}
\end{table}

The Automotive Safety Integrity Level (ASIL) is a combined classification of severity of potential harm, the exposure of the driving scenario and the controllability of the hazardous event~\cite[Part 3]{ISO_26262_2001}.
In case of the protective vehicle controllability levels have to be differentiated by the capability of other traffic participants to control the hazardous event, as there is no driver or permanent supervisor available for the vehicle.
The first hazardous event, the protective vehicle entering moving traffic with full left lock, is classified as ASIL~D because
\vspace{-0.05em}
\begin{itemize}
	\item the hazardous event can cause a fatal crash with multiple cars and high relative velocities (severity S3),
	\item the driving scenario is the most common for this vehicle (exposure E4), and
	\item a fast crossing protective vehicle is hard to control for other traffic participants (controllability C3).
\end{itemize}
\vspace{-0.05em}
This hazardous event can be mitigated by limiting steering left lock in Follow Mode to \SI{3}{\degree} as stated in the first safety goal.
A limited maximal steering angle leads to a reduced lateral velocity.
When the protective vehicle is entering moving traffic with reduced lateral velocity, the hazardous event can be classified as ASIL B due to better controllability for other traffic participants (controllability C1).
Turns on highways typically provide large radii, such that the protective vehicle can follow all turns with the demanded limited left lock.
For this safety goal an ASIL~D hard- and/or software system can be implemented with manageable effort.
As the final safety concept shall not only mitigate but eliminate system level hazards, the mitigated hazard must be reclassified with the assumption that the limitation is implemented.
To eliminate this hazard, we introduce the safety goal to \emph{always maintain a safe distance to the lane marking of the hard shoulder}.

For the following parts of this contribution we choose this second hazard and the corresponding safety goal to exemplify how to derive further safety requirements and how to use these requirements in online monitoring to maintain a safe operation.

\section{Project Controlling Concurrent Change}
As described above, current automotive systems are developed according to the ISO~26262~standard, which proposes a V-model-like development process.
Following the V-model, the aforementioned functional specification from the aFAS project is located in the upper left part of the "V".
While aFAS is following a rather classic implementation of the V-model (specification, implementation, integration, test), the project CCC aims at facilitating the integration process in the right branch of the V-Model, particularly with regard to the addition of updates at runtime.

Due to the complexity of modern automotive systems consisting of an increasing number of system components, the verification of a complete system is becoming more and more complex and time-consuming. 
Integration tests in particular often show the outcome of hidden dependencies between the requirements for single system components, which makes the integration of updated components even more challenging. 
CCC aims at an automated model-based integration process for safety-critical embedded systems which includes automated analysis of the aforementioned dependencies.
For this purpose, the system is modeled from a variety of viewpoints (e.g.~function, timing, platform) using the corresponding requirements as a basis.
However, as not all side-effects, which occur during operation, can be anticipated in the design process, monitoring mechanisms are instantiated at runtime.

In this paper, we describe skills and abilities as one of the models applied at the functional viewpoint which is used in the project context.
In the context of the aFAS use-case we present a knowledge-based approach for the derivation of performance metrics for abilities.

\section{Concept of skills and abilities}
The basic idea behind the skill and ability based representation is to model the system along the human driving task at a functional level.
Dependencies (edges) between skills and abilities (nodes) are expressed in the form of a directed acyclic graph.
The main concept and its origins have been described by \citet{reschka_ability_2015}.
Further literature on the need for an internal representation of the system's performance can also be found in \cite{schlatow2017}.

This can yield a representation of what the system must do how well for fulfilling its functional specification.
In this context we would like to correct a translation error (reversed terms ability and skill) which occurred in~\cite{reschka_ability_2015}, originating from a misinterpretation of a German source~\cite{funke_dorsch_2014}.
For correct translation, we refer the definition of skills and abilities for technical systems given by~\citet{Reschka2016}:
\begin{definition}
A \emph{skill} describes an activity of a technical system which has to be executed to fulfill the defined goals of the system (\emph{translated from German}).
\end{definition}
\begin{definition}
An \emph{ability} describes the quality level of an activity dependent on internal properties and the current operational situation of the system (\emph{translated from German}).
\end{definition}

The differentiation between skills and abilities allows to apply the concept to the development phase (skill graph, Section~\ref{sec:development}) and the use phase (ability graph, Section~\ref{sec:online}) of the system (nomenclature of phases according to~\cite[Chap. 3]{avizienis_basic_2004}.
In this context, the skill graph can be used to derive safety and technical requirements.
The requirements can then be attributed to skills in the skill graph as annotations of the outgoing edges of a skill.
For instantiating the ability graph at runtime, expressive metrics for performance monitoring are required.
In combination with the requirements derived for the skill graph, these metrics yield the basis for determining a performance level for each ability node in the graph at runtime.
The performance level of an ability is represented by the edge weight of the incoming edge and consists of the performance levels annotated to the outgoing dependency edges.

The following sections will present possible applications of skills and abilities in the development and use phase in more detail.
If not mentioned otherwise, we use terms defined by~\citet{avizienis_basic_2004} and focus on the \emph{safety} aspect in the given taxonomy of \emph{dependability}.
For the derivation of safety requirements, we draw the system boundary around the vehicle.
In consequence we propose an extension of the terms given in~\cite{avizienis_basic_2004}: ~\citeauthor{avizienis_basic_2004} define \emph{behavior} as a sequence of \emph{states}.
The external state of a system is defined as "the part of the provider's total state that is perceivable at the service interface" \citep[p. 3]{avizienis_basic_2004}.
Based on this, we differentiate \emph{internal} and \emph{external} behavior, as a sequence of \emph{internal} and \emph{external} states, respectively.
As a result, the \emph{internal behavior} describes what the system does internally to fulfill its function, while the \emph{external behavior} is the behavior which directly influences other traffic participants and can thus cause hazards.

\section{Skills in the development process}
\label{sec:development}
\begin{figure}[htb]
	\vspace{-0.75em}%
	\centering
	\includegraphics[width=0.43\textwidth]{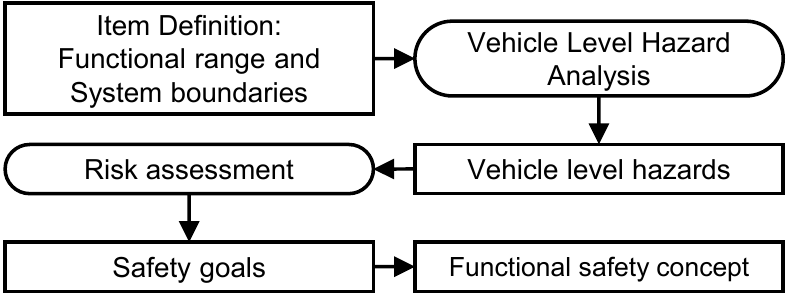}    
	\caption{Top-down development process} 
	\label{fig:conceptphase}
	\vspace{-0.75em}%
\end{figure}

%
The ISO~26262~standard demands the formulation of functional descriptions and concepts for a functional safety concept without giving details of technical solutions. 
With this demand, the resulting concepts can be reused in further stages of development and assessed by external reviewers without struggling with a flood of information.
Fig.~\ref{fig:conceptphase} shows a shortened overview over steps in the concept phase.
\begin{figure*}[bt]
	\centering
	\includegraphics[width=.9\textwidth]{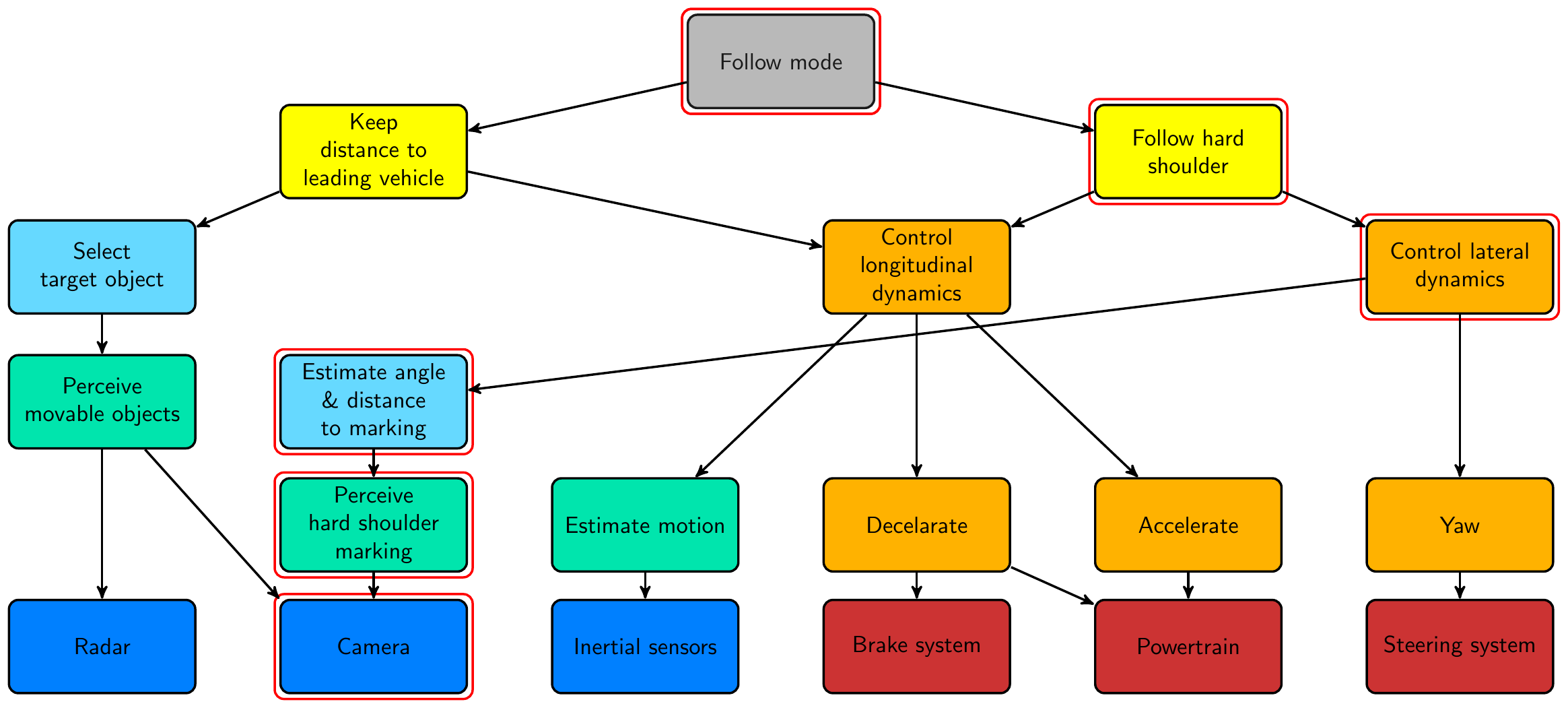}    
	\vspace{-0.6em}
	\caption{Skills for follow mode with categories main (grey), observable external behavior (yellow), perception (green), planning (light blue), action (orange), sensors (blue), actuators (red). Examplatory path for structured derivation of performance metrics from safety goals is depicted by red frames.} 
	\label{fig:skillgraph}
	\vspace{-1.5em}
\end{figure*}

Reschka~\cite{Reschka2016} proposes to use skill graphs in the \emph{item definition} which is the first step in the development phase.
After creating a use case description, the level of automation is determined.
Followed by this, operational scenarios, the intended behavior, and safe states for each scenario must be described.
To cluster and describe operational scenarios of the vehicle, maneuvers can be used to express the behavior of vehicles from an external point of view.

Reschka also proposes a list of driving maneuvers to cover possible traffic scenarios based on literature review and further refinements towards automated vehicles.
Each maneuver can be expressed by physical measures (vehicle A going faster than vehicle B) and semantic relations (lane change from right to left).
Therefore maneuvers describe the \emph{external behavior} of the vehicle.
Which maneuvers have to be fulfilled by the vehicle can be derived from the intended use case and the operational scenarios.
The chosen maneuvers can then be modeled using skill graphs.
This model shall be independent from specific situations which the vehicle could encounter at run-time. 
Skill graphs can provide information how a system is able to fulfill the driving task and which dependencies exist between the system's skills in a functional manner.
In this context, external behavior is characterized by skills which are in direct relation to the described top-level function (cf.~Fig.~\ref{fig:skillgraph}).
The majority of skills in the skill graph describe the \emph{internal behavior} of the system.

Bagschik et al.~\cite{bagschik_identification_2016} used skill graphs combined with computer-aided generation of operational scenarios to derive potentially hazardous scenarios according to the designed use case.
This analysis was conducted without any technical details of implementation.
One resulting skill graph was designed for the maneuver of following the hard shoulder in the state \emph{Follow Mode}.
Each skill from the graphs in the item definition was assigned to the category \emph{main}, \emph{sense}, \emph{plan} or \emph{act} (cf.~Fig.~\ref{fig:skillgraph}).


As the ISO~26262~standard formulates that the HARA ``is based on the item's functional behavior; therefore, the detailed design of the item does not necessarily need to be known.'' \cite[Chap. 3.72]{ISO_26262_2001}
We further define that the mentioned functional behavior is the \emph{external behavior} of the item.

\begin{figure}[hbt]
	\vspace{-0.8em}%
	\centering
	\includegraphics[width=0.4\textwidth]{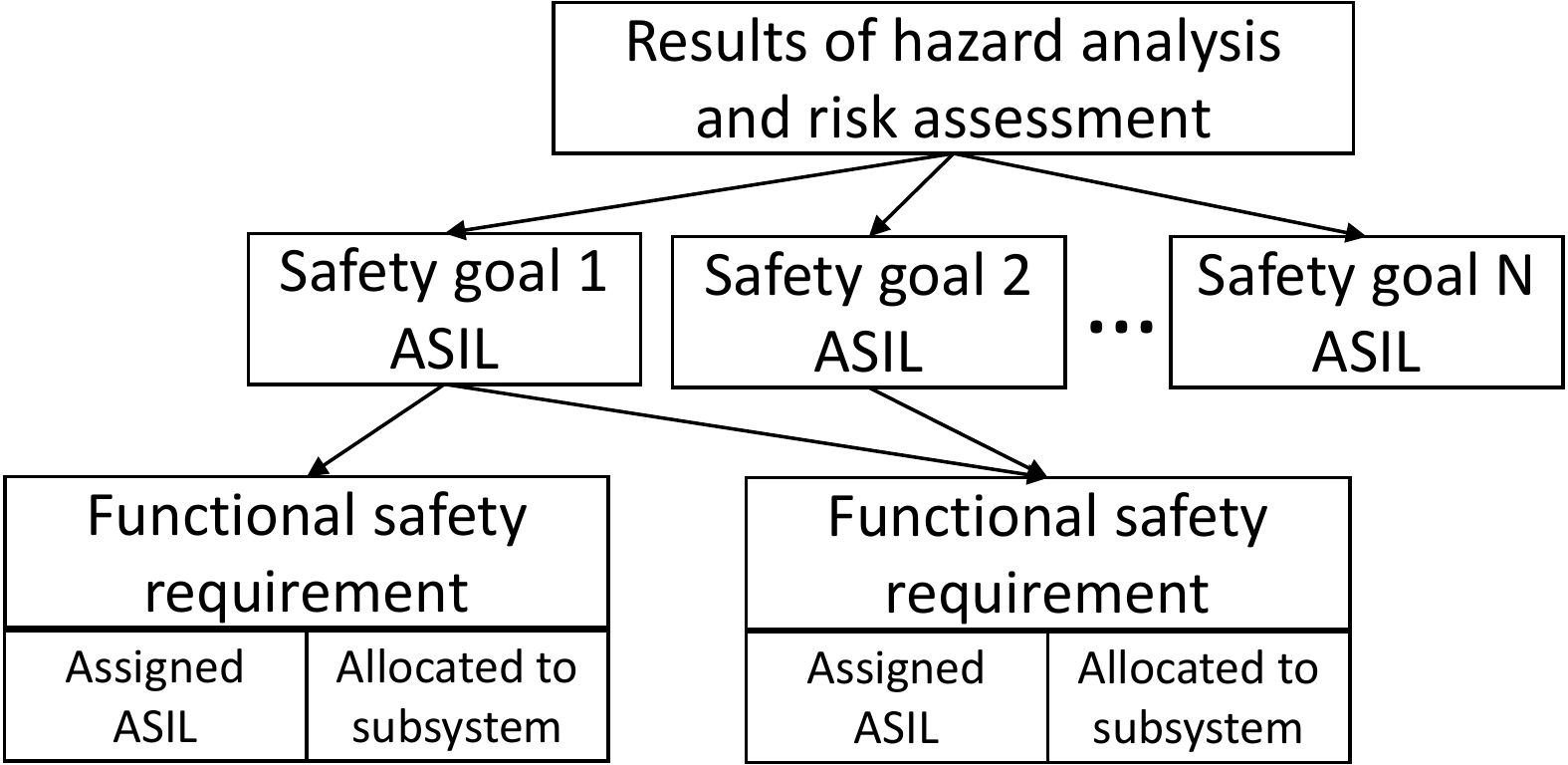}    
	\caption{Relation of HARA, safety goals and functional safety requirements according to \cite{ISO_26262_2001}.} 
	\vspace{-0.5em}
	\label{fig:safetygoals}
	\vspace{-0.9em}%
\end{figure}

The main item for assessment is called AFA-logic as described by \citet{stolte2017} and describes the service provided by the system on the top-level.
Thus, we have to identify which external system states are hazardous in each possible operating scenario of the item.
As mentioned above, the identification of the resulting hazardous scenarios which result from the \emph{external} behavior of the system, yields the basis for safety goals for the \emph{external behavior}.
This directly relates the list of functional safety requirements (cf.~Fig~\ref{fig:safetygoals}) to the externally visible behavioral skills (cf.~Fig~\ref{fig:skillgraph}). 
For the underlying skills, more specific technical requirements can be derived in order to fulfill those functional safety requirements.

The given safety goal \emph{always maintain a safe distance to the left lane marking} would thus entail an exemplary functional safety requirement of the form \emph{the vehicle must always keep a distance of at least \SI{0.28}{\meter} from the left lane marking when following the hard shoulder} for the skill \emph{follow hard shoulder}.
For the underlying skill \emph{control lateral dynamics}, this would translate to a control-performance metric of the form: \emph{the lateral controller must guarantee a maximal overshoot of \SI{0.28}{\meter}}.
These requirements must be met during operation of the vehicle for guaranteeing safe operation.
A concept for monitoring (safety-related) system performance will be presented in the next section.
%

\section{Abilities for online performance monitoring}
\label{sec:online}
For runtime monitoring of the system's performance, the skill graphs derived in the development process can be transformed into ability graphs.
The challenge for monitoring performance in specific situations at run-time is to find expressive metrics even if metrics from different domains are combined for calculating a level of performance.

Abilities can consist of multiple interacting software components which makes them a (sub-)system, according to \cite{avizienis_basic_2004}.
As performance is measured at the interfaces of these sub-systems, the performance attributed to abilities can be defined as \emph{quality of service} (\emph{service} cf.~\cite[p. 3]{avizienis_basic_2004}).

In the field of (autonomic) grid computing, substantial research effort has been spent in recent years to create structured approaches towards a meaningful computation of quality of service in heterogeneous computing systems.
For this purpose, knowledge-based approaches have been developed which apply ontologies (formally ordered representations of terms, obects and their relations) as a basis for knowledge representation (cf.~\cite{sorathia2010} for an overview).
Possible applications range from a specification of quality of service during design- and runtime to reasoning about the quality of composed services \cite{dobson2005}.
These applications are in line with the development process for self-aware automated vehicles which has been proposed in the sections above.

For our application, ontologies provide means to model performance measures for each skill and incorporate identified safety goals and resulting requirements for the abilities.
Additionally, an ontology extending the semantics of the ability graphs allows to define mappings between the vehicle's abilities and software components implementing the respective ability.
With this connection it is possible to explicitly define redundancies within the system to support degradation strategies. 
As ontologies provide a formal description of entities and their relations, they also enable reasoning about sensible combinations of entities at runtime.
Considering reconfigurable systems, as in the CCC project, the relations can be used as a basis to synthesize redundant (software-)sub-systems and corresponding quality metrics at runtime.

However, simply propagating metrics from lower levels unchanged through the graph adds little to no value to the monitoring on higher levels. 
Instead, we aggregate lower level metrics to composed metrics on higher levels which allows us to add a weight to each metric and prioritize certain aspects over others during monitoring. 
During composition, it must be ensured that metrics remain expressive for effective monitoring.
To ensure this, we equip each metric with a history of the metrics it was composed of and utilize this history for the composition of metrics on higher levels of the graph.

Within the ontology, we differentiate between abilities and metrics as shown in Fig.~\ref{fig:performancemetric} a), although slightly simplified for more clarity. 
An ability depicts a node in the ability graph and can depend on other abilities within the graph which is specified by a dependency relation.
A dependency relation is required for each ability to reflect the edges of the graph as shown in Fig.~\ref{fig:skillgraph}.
It can be described by its type, i.e. whether the ability node is the main node, a sub node, a data sink, or a data source.
A category can be assigned in analogy to the skill graph, i.e. by grouping the nodes into visible external behavior, perception, planning, action, sensors, and actuators. 
Each ability has a performance level, and a dedicated mapping regarding the software component(s) which implement this ability and thus provide data for associated performance metrics.
An ability can have one or several metrics from which the performance level is determined.
Each metric can be grouped by whether it is a composed metric or an atomic metric. 
In addition to their composition history composed  metrics are attributed with information about their composition rules in order to avoid fusion of correlated metrics. 
Still, finding appropriate and meaningful metrics for each ability and graph level, aggregation measures, as well as models for performance propagation poses a challenging task and is subject to ongoing research.
\begin{figure*}[htb]
		\centering
		\scalebox{0.63}{%
			\tikzstyle{myblock} = [rectangle, rounded corners=1pt]%
\definecolor{shadedGray}{RGB}{30,30,30}
\begin{tikzpicture}[>=latex]

  \node [draw,ellipse,myblock, shadedGray] (PerformanceMetric) at (107.6838em,59.745em) {PerformanceMetric};

  	\node [draw,ellipse,myblock, shadedGray] (Overshoot) at (97.7261em,56.8998em) {Overshoot};
  	\node [draw,ellipse,myblock, shadedGray] (RiseTime) at (118.4956em,52.7746em) {RiseTime};
  	\node [draw,ellipse,myblock, shadedGray] (ControlError) at (107.6838em,50.0722em) {ControlError};
  	\node [draw,ellipse,myblock, shadedGray] (SettlingTime) at (117.9262em,57.896em) {SettlingTime};

  \node [draw,ellipse,myblock, shadedGray] (Value) at (101.9936em,64.0124em) {Value};
  \node [draw,ellipse,myblock, shadedGray] (Unit) at (114.7964em,64.0124em) {Unit};

  	\node [draw,ellipse,myblock, shadedGray] (ControlMetric) at (107.6838em,54.055em) {ControlMetric};

  	\node [draw,ellipse,myblock, shadedGray] (ControlSurface) at (117.4992em,42.6749em) {ControlSurface};

  	\node [draw,ellipse,myblock, shadedGray] (IntegratedControlError) at (113.3738em,46.3736em) {IntegratedControlError};
  	\node [draw,ellipse,myblock, shadedGray] (SteadyStateOffset) at (97.7261em,46.9426em) {SteadyStateOffset};
  	\node [draw,ellipse,myblock, shadedGray] (AbsoluteControlError) at (101.9936em,42.6749em) {AbsoluteControlError};

  \draw [dashed,->] (PerformanceMetric) to [out=80,in=260,loop,looseness=1]  node [fill=white] {\small hasUnit} (Unit);
  \draw [dashed,<-] (ControlMetric) to [out=10,in=-170,loop,looseness=1]  node [fill=white] {\small isA} (SettlingTime);
  \draw [dashed,<-] (ControlError) to [out=-170,in=30,loop,looseness=1]  node [fill=white] {\small isA} (SteadyStateOffset);
  \draw [dashed,<-] (ControlMetric) -- node [fill=white] {\small isA} (RiseTime);
  \draw [dashed,<-] (PerformanceMetric) -- node [fill=white] { \small isA} (ControlMetric);
  \draw [dashed,->] (PerformanceMetric) to [out=100,in=280,loop,looseness=1] node [fill=white] { \small hasValue} (Value);
  \draw [dashed,<-] (ControlError) -- node [fill=white] { \small isA} (IntegratedControlError);
  \draw [dashed,<-] (IntegratedControlError) -- node [fill=white] { \small isA} (ControlSurface);
  \draw [dashed,<-] (ControlMetric) -- node [fill=white] { \small isA} (ControlError);
  \draw [dashed,<-] (ControlError) -- node [fill=white] {\small isA} (AbsoluteControlError);
  \draw [dashed,<-] (ControlMetric) -- node [fill=white] {\small isA} (Overshoot);

\begin{scope}
  \node [draw,ellipse,myblock, shadedGray] (Ability) at (52.6325em,54.055em) {Ability};
  \node [draw,ellipse,myblock, shadedGray] (SWComponent) at (46.9425em,61.1675em) {SWComponent};
  \node [draw,ellipse,myblock, shadedGray] (AbilityCategory) at (56.9em,44.0975em) {AbilityCategory};
  \node [draw,ellipse,myblock, shadedGray] (AbilityType) at (45.52em,46.9425em) {AbilityType};
  \node [draw,ellipse,myblock, shadedGray] (PerformanceLevel) at (59.745em,62.59em) {PerformanceLevel};
  \node [draw,ellipse,myblock, shadedGray] (String) at (62.59em,51.21em) {String};
  \node [draw,ellipse,myblock, shadedGray] (PerformanceMetric1) at (76.815em,54.055em) {PerformanceMetric};
  \node [draw,ellipse,myblock, shadedGray] (MetricHistory) at (82.9275em,59.1675em) {MetricHistory};
  \node [draw,ellipse,myblock, shadedGray] (MetricType) at (75.3925em,48.365em) {MetricType};
  \node [draw,ellipse,myblock, shadedGray] (MetricCategory) at (72.5475em,61.1675em) {MetricCategory};
  \node [draw,ellipse,myblock, shadedGray] (ComposedMetric) at (85.35em,45.52em) {ComposedMetric};
  \node [draw,ellipse,myblock, shadedGray] (AtomicMetric) at (71.125em,44.0975em) {AtomicMetric};
  \end{scope}
  \draw [dashed,->] (Ability) -- node[below,fill=white] {\small hasAbilityCategory} (AbilityCategory);
  \draw [dashed,->] (ComposedMetric) to [out=80,in=-60,loop,looseness=1] node[xshift=0.5cm,yshift=0.5cm,fill=white] {\small hasMetricHistory} (MetricHistory);
  \draw [dashed,->] (Ability) to [out=20,in=170,loop,looseness=1] node[fill=white] {\small hasPerformanceMetric} (PerformanceMetric1);
  \draw [dashed,->] (Ability) to [out=45,in=-150,loop,looseness=1] node[xshift=0.5cm,fill=white] {\small hasPerformanceLevel} (PerformanceLevel);
  \draw [dashed,->] (Ability) -- node[xshift=-0.2cm,fill=white] {\small hasAbilityType} (AbilityType);
  \draw [dashed,->] (Ability) to [out=120,in=-50,loop,looseness=1] node[xshift=-0.85cm,fill=white] {\small hasSWComponent} (SWComponent);
  \draw [dashed,<-] (MetricType) to [out=-130,in=30,loop,looseness=1] node[xshift=-0.3cm,yshift=0.1cm,fill=white] {\small isA} (AtomicMetric);
  \draw [dashed,->] (PerformanceMetric1) -- node[fill=white] {\small hasMetricType} (MetricType);
  \draw [dashed,<-] (MetricType) to [out=-0,in=145,loop,looseness=1] node[fill=white] {\small isA} (ComposedMetric);
  \draw [dashed,->] (PerformanceMetric1) edge [bend left] node[above,fill=white] {\small hasMetricHistory} (MetricCategory);
  \draw [dashed,->] (ComposedMetric) edge [bend right] node[xshift=0.1cm,fill=white] {\small isComposedFrom} (PerformanceMetric1);
  \draw [dashed,->] (PerformanceMetric1) to node[yshift=0.1cm,xshift=-0.1cm,fill=white] {\small hasName} (String);
  \draw [dashed,->] (Ability) -- node[yshift=0.1cm,fill=white] {\small hasName} (String);
   \draw [dashed,->] (Ability.north west) to [out=140,in=200,loop,looseness=4.2] node[left,xshift=0.5cm,fill=white] {\small hasDependencyOn} (Ability.south west);

\node[align=left, font=\large] at (24,13.8) {Classes and object properties for an ontology describing\\ abilities and corresponding performance metrics };
\node[font=\large] at (18.55,14.05) {a) };
\node[align=left, font=\large] at (38,13.8) {Exemplary ontology for metrics addressing\\control quality};
\node[font=\large] at (33.55,14.05) {b) };
\end{tikzpicture}%
		}
		\vspace{-0.8em}%
		\caption{Ontologies for structured derivation of performance metrics}%
		\label{fig:performancemetric}%
		\vspace{-1.5em}%
\end{figure*}
From the safety goals identified in section~\ref{ref:afas}, we derive metrics for the main ability \emph{Follow mode} and the visible behavior ability \emph{Follow hard shoulder}. 
On the highest level \emph{Follow mode} the identified safety goal of \emph{Do not drive into traffic} must be adhered to at all times.
Thus, it is essential to monitor whether the vehicle enters or is about to enter traffic of the rightmost lane of the motorway. 
Based on this, the safety goal for the ability \emph{Follow hard shoulder} consists of maintaining a safe distance to the lane markings of the hard shoulder at all times. 
To derive monitoring metrics from this safety goal we need to define what constitutes a minimal safe distance.
Thus, it must be evaluated whether the current distance to the lane markings lies within a specified range below the safe distance of \SI{0.28}{\m}.

In our case, the requirements have been defined in the item definition as stated in section~\ref{sec:development}.
The technical safety requirement for \emph{control of the lateral verhicle dynamics} implies the need for monitoring the overshoot of the control algorithms and check whether it stays within the predefined range.
As the control algorithms are dependent on the \emph{estimated angle and distance to the perceived lane markings}, these estimations must provide some measure of uncertainty, e.g. the variance of the estimation. 
The obvious metric derived from this safety goal is to monitor whether the variance falls within a specified range to stay within the bounds of the hard shoulder.
It should also be known whether the variance estimation is optimistic or pessimistic.
Considering the \emph{perception of the lane markings}, we require a representation of invalid results meaning a plausibility check must be performed to determine validity of the detected markings and to detect when existing markings were not detected.

Fig.~\ref{fig:performancemetric} b) shows parts of a performance metric ontology which can be used for monitoring the control algorithms within the graph.
This part of the ontology can be seen as a generalization and further specification of the class hierarchy presented in \cite{maurer2000a}.
In our example, this set of available performance metrics would be applied to monitor control of the lateral dynamics of the vehicle.
A meaningful performance level can e.g.~be calculated by comparing the actual steady state offset during operation with the maximal allowed steady state offset from the corresponding requirement (\SI{0.28}{\meter}).

Summarizing, the proposed ontology provides means for transforming skill graphs into ability graphs for runtime performance monitoring by providing additional semantics for metrics and performance.
The approach of deriving metrics and thresholds from identified safety goals provides a useful method to aid in this process.

\section{Conclusion}
We have outlined how skill and ability graphs can provide a valuable method for modeling the vehicle's capabilities in the development process while also enabling online monitoring.
Employing skill graphs in the item definition allows the derivation of metrics for runtime monitoring directly from the safety goals identified in the HARA.
Thus, since safety goals are defined for the externally visible vehicle behavior, we distinguish external from internal system behavior.

Although ontologies can provide means for derivation and synthesis of performance metrics, it must be mentioned that their power must not be overstated (cf.~\cite{dobson2005}).
They do not provide ready-made solutions, but should be considered as a tool for coming to solutions in the task of introducing expressive monitoring mechanisms from a functional viewpoint.

\section*{Acknowledgment}
This work is part of the DFG Research Unit Controlling Concurrent Change, funding number FOR 1800.

The project aFAS is partially funded by the German Federal Ministry of Economics and Technology (BMWi). 
The consortium consists of MAN (consortium leader), ZF TRW, WABCO, Bosch Automotive Steering, TU Braunschweig, Hochschule Karlsruhe, Hessen Mobil - Road and Traffic Management, and BASt - Federal Highway Research Institute. 

We would like to thank our partners and colleagues in both projects for valuable discussions, Mischa M\"ostl in particular for providing additional input from a different domain.



\renewcommand*{\bibfont}{\footnotesize} 
\printbibliography 

%



\end{document}